# Scalability of data binding in ASP.NET web applications

Toni Stojanovski, Member, IEEE, Ivan Velinov, and Marko Vučković

toni.stojanovski@eurm.edu.mk; vuckovikmarko@hotmail.com

*Abstract*—*ASP.NET web applications typically employ server controls to provide dynamic web pages, and data-bound server controls to display and maintain database data. Most developers use default properties of ASP.NET server controls when developing web applications, which allows for rapid development of workable applications. However, creating a high-performance, multi-user, and scalable web application requires enhancement of server controls using custom-made code. In this empirical study we evaluate the impact of various technical approaches for paging and sorting functionality in data-driven ASP.NET web applications: automatic data paging and sorting in web server controls on web server; paging and sorting on database server; indexed and non-indexed database columns; clustered vs. non-clustered indices. We observed significant performance improvements when custom paging based on SQL stored procedure and clustered index is used.*

## I. INTRODUCTION

In the last decade we are observing increased use of web applications. This is a consequence of many factors: zero-client installation, server-only deployment, powerful development tools, growing user base etc. Furthermore, competition and the quickly changing and growing user requirements create a demand for rapid development of web applications. Microsoft Visual Studio (MVS) is the dominant web applications development environment of today. MVS provides numerous mechanisms to support rapid development of ASP.NET applications. Most developers tend to use the default settings for the ASP.NET server controls which are arguably the most significant enabler of the rapid development. Though the ASP.NET server controls can significantly decrease the application's "time to market", at the same time they can reduce performance and scalability of the web application. Analysis of factors which influence the response time of web applications is an active area of research [1]. In this paper, we demonstrate the importance of adding custom program logic to the data binding mechanism of ASP.NET server controls, that is, the mechanisms used to maintain and display data, in order to improve performance and scalability of web applications.

Here we address the following research questions: (i) Can custom stored procedures for fetching, sorting and paging the results provide better response time and improved scalability compared to the automatic data binding of ASP.NET server controls? (ii) What is the impact of indices on response time when sorting and paging the results? (iii) What is the dependence of the response time on the number of database records?

The outline of our paper is as follows. In Section II we explain the basics of data binding, paging and sorting in ASP.NET applications. Section III describes our test environment and the testing approach. Test environment is used to measure the response time for ASP.NET pages which implement various methods for data fetching and display. In Section IV we explain the results from the tests. Section V concludes the paper and outlines further research.

## II. DATA BINDING IN ASP.NET APPLICATION

When using ASP.NET data-bound control such as GridView to display data from a database, the simplest way is to bind the data-bound control with a data-source control, which connects to the database and executes a query. When using this scenario, the data-source control automatically gets the data from the database server [2] after the Page.PreRender event in the page life cycle [3], and displays it in the data-bound control. This is the code that is used for the data-source control to bind with the database.

```
<asp:SqlDataSource ID="SqlDS1" runat="server"
ConnectionString="<%$ ConnectionStrings:tdbConn %>"
SelectCommand="usp_autoDataBinding"
SelectCommandType="usp_autoDataBinding"/>
```

Following code connects a GridView control with the data-source control.

```
<asp:GridView ID="GridView1" DataSourceID="SqlDS1" ...>
<Columns>
<asp:BoundField DataField="ID" HeaderText="ID"
SortExpression="ID" .../>...
```

Following stored procedure is used to query data from the database

```
CREATE PROCEDURE [dbo].[usp_autoDataBinding] AS
BEGIN
    SELECT * FROM testTable
END
```

Code 1. Query that returns all data from a database.

When there are many records to display in a web page, it is a common practice to show only a limited number of records (a page of records) and to allow the user to navigate through the pages of records i.e. to use "data paging". Data-bound controls such as GridView have a built-in mechanism for sorting and paging [2]. First, the data-source control gets *all* the data from the database (see Code 1) in a dataset, and then the ASP.NET data-bound control is responsible to sort the dataset and display only a small number of records enough to fill a page. For example, a dataset can contain millions of records, but a web page displays only 10 of these records. This approach poses two problems: (i) lots of data is transferred between the database server and the web server, which is especially an issue in a multi-server deployment scenario; (ii) there is a significant consumption of CPU and memory resources to sort large datasets on the web server. Clearly, these problems have significant negative impact on the performance and scalability of the application.

We expect that these problems can be reduced if one uses a custom SQL stored procedure which sorts and returns only the records that will be displayed in the web page. Thus, the network consumption can be reduced, and the database server gets the responsibility to sort and page the records. We are using the following implementation of a stored procedure to page and sort the results:

```
CREATE PROCEDURE [dbo].[usp_selectGridViewOrderByID]
@pageNumber int, @PageSize int = 10
AS
DECLARE @Ignore int
DECLARE @LastID int
IF @pageNumber > 1
BEGIN
    SET @Ignore = @PageSize * @pageNumber
    SET ROWCOUNT @Ignore
SELECT @LastID = ID from testTable ORDER BY ID ASC
END
ELSE
BEGIN
    SET @LastID = 0
END
SET ROWCOUNT @PageSize
SELECT * FROM testTable WHERE ID > @LastID ORDER BY ID
ASC
```

Code 2. SQL stored procedure which supports custom data sorting and paging.

This stored procedure logically divides the records from table testTable into pages of size @pageSize records, and returns the records from page @pageNumber. Records are ordered by field ID.

Performance of this stored procedure greatly depends on the use of index on field ID and the type of index: clustered or non-clustered [4]. By using indexed data structure we can significantly improve the time required for getting the data out of the database.

In our test environment, we tested several scenarios which differ in the following parameters: (i) Number of records in database; (ii) Use of clustered and non-clustered database indices; (iii) Automatic data paging and sorting in ASP.NET server controls vs. paging and sorting in SQL stored procedures

## III. TESTING APPROACH

For our test environment we used HP 550 Notebook with following characteristics: Processor: Intel Core 2 Duo CPU T5470 @1.60 GHz; RAM: 2.00 GB; OS: Windows 7 Professional 32–bit; Internet Information Server (IIS) Version 7.5.7600.16385; Visual Studio 2010 Ultimate; SQL Server 2008 R2.

For the test environment we created a web application with two web pages – one for each of the data binding, paging and sorting approaches as explained in Section II. First page uses Automatic Data Binding (ADB): its GridView control has paging and sorting allowed, and is populated with a stored procedure that gets all the records from the database as in Code 1. Second page uses a custom stored procedure (see Code 2) to query the results populate its GridView control. The stored procedure orders the results at the SQL server, and returns only the records that will be shown in the web page. Sorting field and page number are passed to the web page in the query string of the HTTP request.

We are interested in the time required to process a HTTP request on the web server. We used ASP.NET tracing feature to determine when the data binding occurs in the page lifecycle. We start the timer at Page_Init event and end the timer at Page_SaveStateComplete event, which is after the Page_PreRender event. Functionality in the test pages is kept to minimum in order to avoid the impact of other factors on the response time. Web pages are responsible to record the response time in a text file, and these time measurements are later analyzed.

Our test database has one table with five fields. Records in the table are populated with random values.

TABLE 1: FIELD IN TEST TABLE TESTTABLE.

| Name | Type |
|------|------|
| ID | Int, autoincrement |
| TextField | Varchar(50) |
| IntField | Int |
| BoolField | Bit |
| DateField | Datetime |

## IV. MAIN RESULTS

In Figure 1 we show the results when the data table has 1.000.000 records. For each of the two test web pages there are three types of results, depending on which field was used to sort the results. In this experiment table testTable has no indices. Every web page and the corresponding sorting and paging approach was tested 500 times. Measured response times were grouped into 10 bins. Figure 1 shows the frequency of the bins.

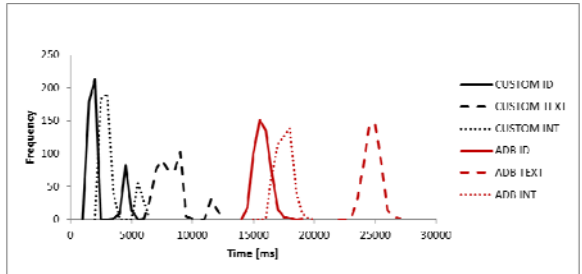

Figure 1. Automatic Data Binding vs. Custom Paging without indices.

Black curves in Figure 1 show the results when the data sorting and paging are executed in the ASP.NET web page using automatic data binding. The problem with this method is that the data-source control needs to fetch 1.000.000 results from the database first before the ASP.NET can sort all these records, and identify the records belonging to the required page. Field ID is auto incremented, and the records are physically sorted by this field in the database. Therefore, the time needed for the ASP.NET to sort the dataset by ID field is faster than the other two sorts. Response time when ordering by TextField and IntField is different because it is faster to sort integer than textual fields.

Red curves in Figure 1 represent the results when the sorting and paging are done in the database server using an SQL stored procedure as in Code 2. Response time is significantly shorter compared with ADB. The reason is twofold: (i) SQL server is optimized for working with large datasets; (ii) SQL stored procedure returns only a small number of records sufficient to fill the ASP.NET web page. The difference in the response time when sorting by different columns is caused by same reasons as explained for ADB.

Next, we repeated the above tests when there are indices in the table testTable. The aim is to see the differences in response time when clustered and non-clustered indices are used. The reader should note that Figure 2 uses a different time scale from Figure 1.

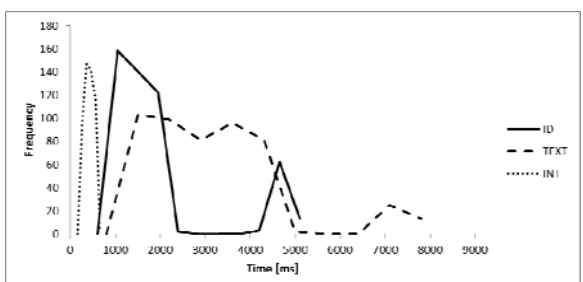

Figure 2. Custom Paging. Clustered index on IntField.

In Figure 2 table testTable is clustered by IntField, and fields ID and TextField have non-clustered indices. Because of the clustered index, the response time when sorting by IntField is significantly smaller compared to the response time when sorting by ID and TextField. This pattern repeats when the clustered index is on a different field i.e. ID or TextField: the response time when sorting by a clustered index is shorter that the response time when sorting by a non-clustered index.

The presence of indices has no impact on the response time when the sorting and paging is done in the ASP.NET web page on the web server using the SQL stored procedure from Code 1. Results are identical to those shown by black curves in Figure 1.

A peculiar property of Figure 1 and Figure 2 is the presence of two peaks. They appear when sorting is done on a non-indexed field (all curves in Figure 1), or a field with a non-clustered index (ID and TextField in Figure 2). This means that there are two groups of time responses for the SQL stored procedure in Code 2. The problem lies in the second select statement "SELECT * FROM testTable WHERE ID > @LastID ORDER BY ID ASC" in Code 2. We detected that the response time is much longer when the input parameter @pageNumber < 18000. @LastID is smaller for smaller values of @pageNumber. Consequently, the SELECT statement sorts and returns a larger data set, and the time needed for its execution increases. The SQL server uses the index file to identify the ordering of records, and then joins the index file with the records from the table testTable, and finally returns every column in the record (note the use of the * sign which means that *all* table columns are returned). As the size of the dataset increases, SQL server uses more memory to sort the dataset. If the dataset consumes more memory than the amount available to the SQL server process, then the SQL server starts to use virtual memory which is much slower than RAM memory. In our test environment, SQL server starts using the virtual memory when the number of records in the dataset is larger than 820,000 (first 18,000 pages with 10 records each are skipped). The above argument holds for both cases - sorting is done on a non-indexed field or a field with a non-clustered index.

However, when the sorting is done by the clustered index field (e.g. IntField in Figure 2), then it can be noticed that there is a single peak since the records in the table are already physically ordered by the clustered index field.

We repeated the measurements for a different number of records in the table: 100.000, 200.000, 500.000, and 1.000.000 records. The aim was to test the dependency of the response time on the number of records. As expected, the response time is larger for larger number of database records. Response time grows faster with the number of records for ASP.NET server sorting and paging compared to SQL server sorting and paging. Figure 3 and Figure 4 show the relation between response times averaged over 500 tests and the number of records in the table testTable. The fastest response time and the slowest growth with the number of records in table testTable is achieved when using an custom paging with a clustered index, followed by custom paging with a non-clustered index, followed by custom paging without index, followed by web server sorting and paging.

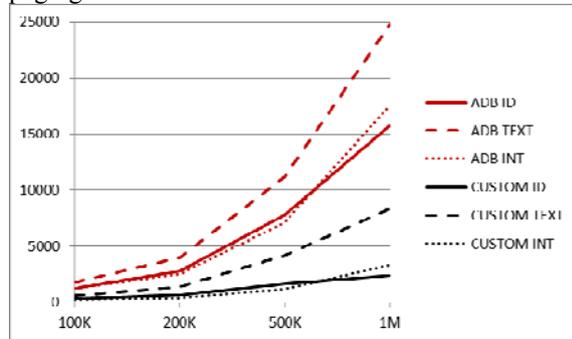

Figure 3. Average response time [ms] vs. number of table records when using Automatic Data Binding (ADB), and custom paging without indices (CUSTOM).

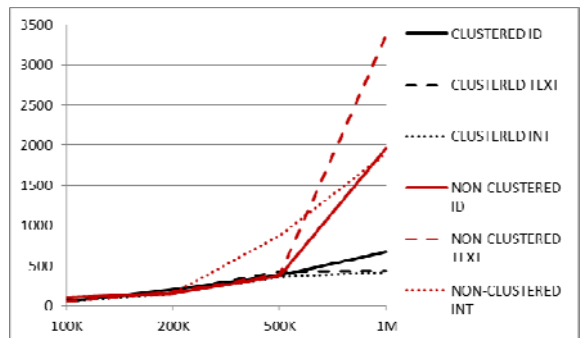

Figure 4. Average response time [ms] vs. number of table records when using custom paging and sorting by non-clustered index and clustered index

Above mentioned tests were repeated in a distributed deployment scenario: MS SQL server was installed on one physical server, and IIS web server was installed on another server. We observed dramatic increases in the response time depending on the network speed for the test cases where the sorting and paging is done on the web server and consequently significant amounts of data travel over the network.

### A. Improved SQL sorting and paging

The problem with the second select statement from Code 2 mentioned before in Section IV can be solved by this modification of the stored procedure:

```
CREATE TABLE #t(x INT)
SET ROWCOUNT @PageSize
INSERT INTO #t
SELECT [int] FROM testTable WHERE [int] > @LastID ORDER BY [int] ASC
SELECT testTable.* FROM #t
LEFT JOIN testTable ON #t.[x] = testTable.[int]
```

Code 3. Modifications to stored procedure from Code 2.

"SELECT *" statement from Code 2 is broken into two parts: First part orders the records by the indexed field and stores only the indexed field into a temporary table #t. Only @PageSize records are stored. No join is done between the index and the records in the table testTable, and thus the execution time is very short for the first part. Second part joins the records from the temporary table #t with the records from the original table testTable. Since the join is done on @PageSize records only (e.g. 10 records), the second part finishes very quickly too.

Figure 5 demonstrates orders of magnitude improvement when the modified SQL stored procedure from Code 3 is used. Similar improvement is achieved when the sorting and paging is done on a field with non-clustered index.

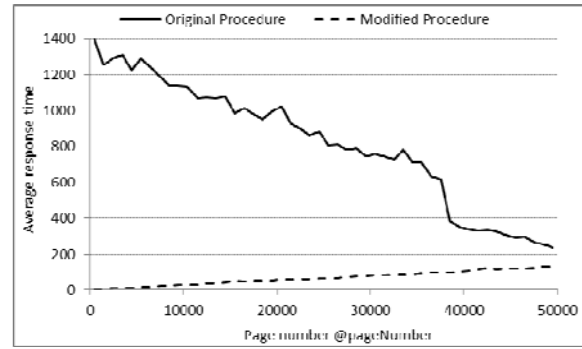

Figure 5. Average response time [ms] vs. page number @pagenumber when sorting by clustered indices.

### B. Scalability

Scalability of the data binding mechanism is tested via Microsoft Test Manager [5]. This testing tool can simulate virtual users that request web pages from the IIS simultaneously. It allows tracking the system and application performance by monitoring different counters such as available memory, processor usage, average page response time etc. We conducted load tests to determine how our application preformed under increased user load. In our tests, we started with a small number of virtual users, and then slowly increased them up to 140 virtual users. For more than 140 users, our test server runs out of For each number of virtual users, we measured the average page response time. Each load test ran for 12 minutes where the initial 10 minutes were used for increasing the number of virtual users, and the remaining 2 minutes were used as cool down period. During the cool down period, the system finishes the started requests and no new requests are processed [6]. We started with 5 users and every minute we increased the number of users by 15, until we reached the total time of 10 minutes and 140 virtual users. As it can be seen from Figure 6, the modified stored procedure offered shorter response times compared with the original stored procedure. This is valid when sorting by any field. Thus, the scalability of the application was improved.

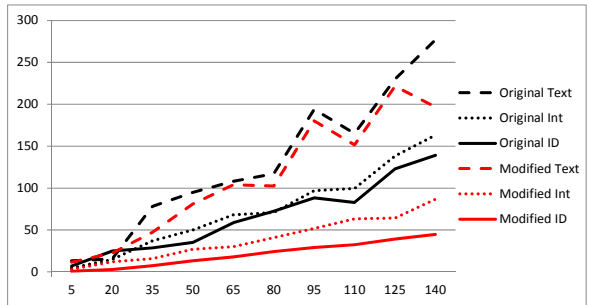

Figure 6. Average response time [s] vs. number of virtual users using original and modified stored procedure.

## V. CONCLUSION

Using the default options of the ASP.NET data bound controls allows for rapid development of sorting and paging functionality. If an ASP.NET data-source control is used to fetch all the data from the database, and then a data-bound control sorts and pages the dataset, then the response time can grow quickly with the size of the returned dataset.

An SQL stored procedures implementing sorting and paging on the SQL server ought to be used when high performance and low consumption of resources are required. It takes less time to fetch the dataset, and then to send to ASP.NET only the records that will be displayed. The response time can be further decreased if the sorting and paging is done on field with indices. Best results are achieved for clustered indices.

Scalability of web applications is significantly improved when paging and sorting is based upon the modified stored procedure from Code 3.